\documentclass[a4paper,twoside,10pt]{article}

\input{jgrg19.sty}

\usepackage{bm}

\begin{document}

\pagestyle{fancy}
\fancyhead{}
  \fancyhead[RO,LE]{\thepage}
  \fancyhead[LO]{Swastik Bhattacharya}                  
  \fancyhead[RE]{Endstates in scalar field collapse}    
\rfoot{}
\cfoot{}
\lfoot{}

\label{P52}

\title{Singular and non-singular endstates in massless scalar field collapse}

\author{Swastik Bhattacharya\footnote{Email address: swastik@tifr.res.in}}
\address{Department of Astronomy and Astrophysics, Tata Institute of Fundamental 
        Research,\\
        Mumbai 400005, India}
\abstract{
We study the collapse of a massless scalar field coupled to gravity. 
A class of blackhole solutions are identified. We also report on a 
class of solutions where collapse starts from a regular spacelike surface but 
then the collapsing scalar field freezes. As a result, in these solutions, a
black hole does not form, neither is there any singularity in the future.
}

\section{Introduction}
The endstate of gravitational collapse is of great interest in gravitational 
physics. The formation of spacetime singularities in gravitational collapse 
and formation of black holes is an issue, which has been investigated in much 
detail in Einstein's theory. The occurrence of singularities indicates the breakdown 
the known laws of physics and offers a regime where the quantum gravity effects 
would be important. As is well known, dynamical evolution of matter fields in a 
spacetime generically yields a singularity, provided reasonable physical conditions 
are satisfied such as the causality, a suitable energy condition ensuring the 
positivity of energy density, and formation of trapped surfaces. This result tells 
us what might happen in those stars, where continual collapse occurs. Therefore it is 
important to investigate whether continual collapse always leads to a singular 
endstate or any other possibility also exists. 

                                     The case of gravitational collapse of a massless scalar field 
is of particular interest in both collapse situations as well as cosmological
scenarios. In cosmology, special importance is attached to the 
evolution of a scalar field, which has attracted a great deal of attention 
in past decades. In gravitational collapse studies, the nature of 
singularity for massless scalar fields has been examined and a number
of numerical and analytical works have been done in recent years on 
spherical collapse models 
from the perspective of the cosmic censorship hypothesis. The issues that have been 
raised above regarding gravitational collapse, can be investigated for massless scalar field
models. This would provide us with some insight into a collapsing matter field in 
general.

In the present study, we outline a mathematical structure to deal with the 
evolution of massless scalar fields in a spherically symmetric  spacetime in 
comoving coordinates. For massless scalar field, this coordinate system breaks 
down when the gradient of the scalar field becomes null. However, as will be shown 
later, in models satisfying some physically reasonable conditions, the gradient of 
the scalar field always remains timelike. So our analysis holds for this class of 
models. In this case, the energy-momentum tensor of the massless scalar field 
has exact correspondence with that of a stiff fluid. So this analysis is 
applicable to stiff fluid collapse also.

Using this formalism, we then examine the classes of collapsing models. In 
particular we find a class of black hole solutions where the future singularity 
is spacelike. We also  point out a class of models where the collapse starts from 
a regular initial surface but with time the rate of collapse decreases so that neither 
any trapped surface nor any singularity is formed in the future.

\section{The Basic Formalism}
In this section, we set up the Einstein equations for the 
massless scalar field in comoving coordinates.

\subsection{Matter field and coordinate system}

The energy-momentum tensor for the massless scalar field is 
\begin{equation}
 T_{ab}=\phi_{;a}\phi_{;b}-\frac{1}{2}g_{ab}(\phi_{;c}\phi_{;d}
g^{cd}).
\end{equation}
In this case, the strong energy condition is always satisfied. We choose the 
comoving coordinate system, in which $T_{\mu \nu}$ and $g_{\mu \nu}$ are diagonal.
The coordinates are $t$, $r$, $\theta$ and $\phi$. This implies that there are only 
two choices: $\phi=\phi(t)$ or $\phi=\phi(r)$. Here we shall consider only the case, 
where $\phi=\phi(t)$. The equation of state is $\rho=P_r=P_\theta$, which is that of 
a stiff fluid. The coordinate system breaks down when $\phi,_\mu$ is null.
 The metric is
\begin{equation}
 ds^2=e^{2\nu} dt^2-e^{2\psi} dr^2-R^2 d\theta^2-R^2 sin^2{\theta} d\phi^2
\end{equation}

\subsection{Einstein equations}

The Einstein equations are
\begin{equation}
 R_{\mu \nu} -\frac{1}{2}R g_{\mu\nu} =- \kappa T_{\mu \nu}
 \end{equation}
We work in such units so that $\kappa$ is one. We define a function $F$ by 
\begin{equation}
 G-H=1-\frac{F}{R}\label{eqn:P52_e2}
\end{equation}
where $G$ and $H$ are given by 
\begin{equation}
 G=e^{-2 \psi} R'^2
\end{equation}
and 
\begin{equation}
 H=e^{-2 \nu} \dot{R}^2 .
\end{equation}

Then we can write down the Einstein equations in the form \cite{P52_psj}
\begin{equation}
\rho=\frac{F'}{R^2 R'},P_r=-\frac{\dot{F}}{R^2 \dot{R}},
\end{equation}
\begin{equation}
 \nu'(\rho+P_r)=2(P_\theta-P_r)\frac{R'}{R}-P_r'\label{eqn:P52_e3},
\end{equation}
\begin{equation}
 -2\dot{R}'+R'\frac{\dot{G}}{G}+\dot{R} \frac{H'}{H}=0\label{eqn:P52_e1}
\end{equation}
where the density $\rho$ is given by 
\begin{equation}
 \rho= \frac{1}{2}e^{-2\nu}\dot{\phi}^2
\end{equation}

\section{Collapse of the scalar field}

The massless scalar field collapses from a regular initial spacelike surface. 
At the initial time $t_i$, the following functions of the comoving radius $r$ are specified.
\begin{equation}
 \nu(t_i,r)=\nu_0(r),\psi(t_i,r)=\psi_0(r),R(t_i,r)=r,\phi(t_i),F(t_i,r)=F_0(r)
\end{equation}
Since we are considering a collapse solution, we discuss only the cases where $\dot{R}\leq0$, 
i.e. the area radius gets smaller with time. For convenience, we define a quantity $v$ by the 
relation $v=\frac{R}{r}$. There would be a strong curvature singularity at $v=0$.

\subsection{Regularity conditions}

Here we consider those solutions that satisfy the following regularity conditions. 
A physically reasonable collapse solution satisfies these conditions. All the metric 
functions are $C_2$ at any non-singular spacetime point. Also $\rho $ , $P_r$ and $ P_\theta$ 
are finite at any regular spacetime point. Further we take $ R'> 0$ for all regular 
spacetime points. This corresponds to the condition that there are no shell-crossings 
in the spacetime. Since $\rvert{\phi,_{\mu}\phi^{,\mu}}\rvert  \varpropto \rho$, as long 
as $\rho>0$, the comoving coordinates can be used. We 
consider here only those cases where $\rho(r,t_2)\leq \rho(r,t_1)$ if $t_2<t_1$.
This means that during the collapse, the density should not decrease with time.
Then if the collapse starts from a regular spacelike surface where the gradient 
of the scalar field is timelike, then throughout the collapse, the gradient remains 
timelike. The F.R.W. solutions are examples of this type.

\subsection{A class of blackhole solutions}

The variables $r$ and $t$ can be interchanged to $r$ and $v$ in all the equations. 
For convenience, we define a new function $M$ by the relation 
\begin{equation}
 F=r^3 M
\end{equation}
Then the system of the Einstein equations can be reduced to the two equations.
\begin{equation}
 v'= -\frac{3M+rM,_r+vM,_v}{2rM,_v} \label{eqn:P52_v'};
\end{equation}
 and 
\begin{equation}
 \dot{v}= - \frac{e^{\nu}}{\sqrt{v}}\sqrt{[\frac{v}{r^2}(G-1)+M]} \label{eqn:P52_vdot}.
\end{equation}
where $e^\nu= \frac{v}{\sqrt{-2M,_v}}$ and $G=-\frac{v^2M,_v(v+rv')^2}{f^2(r)}$ and  $f(r)$ 
is a function of integration.
The last two equations constitute a integrable Pfaffian when $M$ satisfies the equation,
\begin{equation}
\frac{\partial \dot{v}}{\partial r}+ v' \frac{\partial \dot{v}}{\partial v}- \dot{v} \frac{\partial v'}{\partial v}=0 \label{eqn:P52_2pde}
\end{equation}
This is a Monge-Ampere type 2nd order p.d.e. for $M$. However since the expression for 
$\dot{v}$ has square roots in it, sometimes it would be more convenient for us to consider 
the equation,
\begin{equation}
\frac{\partial \dot{v}^2}{\partial r}+ v' \frac{\partial \dot{v}^2}{\partial v}- 2 \dot{v}^2 \frac{\partial v'}{\partial v}=0 \label{eqn:P52_spde}
\end{equation}
The solutions of \eqref{eqn:P52_spde} are solutions of \eqref{eqn:P52_2pde} also; 
except when $\dot{v}=0$.
Now we are in a position to analyze 
the consequences of the Einstein equations.

The time taken for a comoving shell $r$ to reach the singularity is given by 
$t_s(r)= \int_1^0 \frac{1}{\dot{v}}dv$.
It can be shown that the regularity conditions imply that $M$ must be of the form 
$M= \frac{m_0}{v^3}+r^n g(r,v)$, where $n\geq2$. In terms of $M$ and $v$, the density 
can be expressed as $\rho= -\frac{M,_v}{v^2}$. From these relations, it can be shown that 
if $t_s(r)$ is continuous, then $t_s(r)= constant$. Which means that, only spacelike 
singularities are formed. Therefore this would give us a class of black hole solutions \cite{P52_scal}.

\subsection{Collapse without formation of singularity in the future}

Using the formalism that we have presented here so far, we can prove an interesting result \cite{P52_scal}. \\
\textit{If there is a solution which satisfies the regularity conditions, and for which $\dot{v}\leq0$ 
and $v'(r,t)\geq    b$, where $b>0$ for $r_1\leq r\leq r_2$ for some $r_1>0$ and $r_2$, and 
$t\in(t_i,\infty)$, then the comoving shells never become singular.} \\

For these solutions, it can be shown that $\dot{v}<l$ for any $l>0$ for an infinite interval 
of coordinate time. This means that the collapsing matter would freeze eventually. It is important to 
note though that this result does not guarantee the existence of such singularity-free solutions.
To see whether such solutions are there or not we consider \eqref{eqn:P52_2pde} and find out an 
appropriate boundary condition which would result in the freezing of collapse.

To this end, we set the following Cauchy boundary condition for equation \eqref{eqn:P52_2pde} for $M$.
On the surface, $v=\chi(r)$, where $\chi(r)$ is some regular function of $r$, $M(v,r)$ and 
$\frac{\partial M(v,r)}{\partial n}$, i.e. $M$ and its derivative normal to the surface are given.
We set the boundary condition in such a way that both of them are analytic functions. Also 
they are chosen so as to satisfy $[\frac{v}{r^2}(G-1)+M]\mid_{v=\chi(r)}=0$ and $\chi'(r)\neq v'(v,r)\mid_{v=\chi(r)}$.
From this, it can be shown that $[\frac{v}{r^2}(G-1)+M],_v\mid_{v=\chi(r)}=0$. Partial differentiation 
of both the sides of \eqref{eqn:P52_2pde} with respect to $v$, gives us $\frac{\partial^3 M}{\partial^3  v} \mid_{v=\chi(r)}$ in terms of $M(v,r)\mid_{v=\chi(r)}$ and $\frac{\partial M(v,r)}{\partial n}\mid_{v=\chi(r)}$. Then $[\frac{v}{r^2}(G-1)+M],_{vv}\mid_{v=\chi(r)}$ can also be found out. We choose the two given functions 
in such a way so that $[\frac{v}{r^2}(G-1)+M],_{vv}\mid_{v=\chi(r)}>0$. Further they have to satisfy 
\begin{equation}
 [\frac{\partial \dot{v}}{\partial r} +  v' \frac{\partial \dot{v}}{\partial v}]\mid_{v=\chi(r)}=0 \label{eqn:P52_lc}
\end{equation}
\eqref{eqn:P52_lc} is the condition that guarantees that the solution to \eqref{eqn:P52_spde} would also 
be a solution to \eqref{eqn:P52_2pde}. Now using 
Cauchy-Kovalevsky theorem, it can be proved that a solution of \eqref{eqn:P52_spde} exists 
in some neighbourhood of the curve $v=\chi(r)$. The $v=\chi(r)$ curve divides this neighbourhood 
into two parts. In one part, $v$ increases along any line $r=r_2$. For the 
collapse solution, we need to consider only this part. Since this is a solution to 
\eqref{eqn:P52_2pde} as well, such a solution to the Einstein equations exist.

Now we show that this solution does not have a future singularity. For some value of $r=r_1$, 
$\dot{v}$ 
can be written as
\begin{equation}
\dot{v}=-a_1(r_1)(v-v_c)+ O[(v-v_c)^2]
\end{equation}
 where $a_1(r_1)>0$. Now it can be shown that 
the time $t(v_1, r_1)$ taken by the comoving shell $r_1$ to reach $v_1=\chi(r_1)$ is infinite. The proper 
time taken by a comoving shell $r_1$ to reach $v_1$ is $\tau(v_1,r_1)= \int_{t_i}^{t(v_1,r_1)} e^\nu dt$. 
Since $e^\nu(t,r_1)$ has a positive minimum in the range $(t_i, \infty)$, $\tau(v_1,r_1)$ 
is also infinite. So collapse does not lead to a singularity in the future. It can also be 
shown that the proper time for the formation of trapped surface is also infinite. This is 
consistent with the singularity theorems as formation of trapped surfaces implies that singularity 
would form eventually because the strong energy condition is satisfied.

The above analysis tells us that there would be solutions where the scalar field collapses from a 
regular spacelike hypersurface and ultimately the collapse stops. Trapped surfaces do not form, 
neither is there any future singularity. However the present analysis does not tell us whether it would 
be possible to extend these solutions in the past from the regular spacelike hypersurface 
mentioned earlier.

\section{Discussion}

The formalism described here is effective in finding out about the endstate of 
the collapse of massless scalar fields and stiff fluids. Using this formalism, 
it is possible to identify a class of black hole solutions. For massless scalar 
fields, there can be collapse models where collapse ultimately stops. Trapped 
surfaces do not form and the endstate is non-singular. However it is not clear 
whether these solutions can be extended in the past indefinitely. They might be 
of interest as examples of collapse models where collapse slows down, there is 
no bounce, still neither a black hole forms nor is there any 
singularity in the future. If time-reversed, these models might be of interest 
in cosmology as the initial singularity would be absent in that case. It would 
be of interest to see whether it is possible to find out about the endstate of 
collapsing matter proceeding in the way described here with different equations 
of state.

\end{document}